\documentclass[letterpaper,12pt]{amsart}  

\usepackage{fullpage}
\usepackage{amsmath}
\usepackage{amsthm,amsbsy}
\usepackage{mathcomp}
\usepackage{textcomp}

\usepackage{amssymb}
\usepackage{graphicx}
\usepackage{graphicx}
\usepackage{dcolumn}
\usepackage{bm}
\usepackage{amsfonts}
\usepackage{latexsym}
\usepackage{colordvi}
\usepackage{color}
\usepackage{booktabs}
\usepackage{subfig}
\usepackage{graphicx}
\usepackage{stmaryrd}
\input xy
\xyoption{all}

\newcommand{\commentout}[1]{}
\newcommand{\nwc}{\newcommand}
\nwc{\nn}{\nonumber}

\nwc{\nwt}{\newtheorem}
\nwt{cor}{Corollary}
\nwt{proposition}{Proposition}
\nwt{corollary}{Corollary}
\nwt{theorem}{Theorem}
\nwt{summary}{Summary}
\nwt{lemma}{Lemma}
\nwt{definition}{Defintion}
\nwt{remark}{Remark}

\nwc{\FF}{\mathcal{F}}
\nwc{\xx}{\mathbf{x}}
\nwc{\CC}{\mathbb{C}}
\nwc{\ZZ}{\mathbb{Z}}
\nwc{\RR}{\mathbb{R}}
\nwc{\bk}{\mathbf{k}}
\nwc{\bz}{\mathbf{z}}
\nwc{\bom}{\boldsymbol\omega}
\nwc{\bn}{\mathbf{n}}
\nwc{\bN}{\mathbf{N}}
\nwc{\PP}{\mathcal{P}}
\nwc{\PO}{\mathcal{P}_{\rm o}}
\nwc{\POk}{\mathcal{P}_{{\rm o},k}}
\nwc{\QM}{\mathcal{Q}_{\rm m}}
\nwc{\QMb}{\mathcal{\hat Q}_{\rm m}}
\nwc{\QF}{\mathcal{Q}_{\rm f}}
\nwc{\QMk}{\mathcal{Q}_{{\rm m},k}}
\nwc{\QFk}{\mathcal{Q}_{{\rm f},k}}
\nwc{\PF}{\mathcal{P}_{\rm f}}
\nwc{\PFk}{\mathcal{P}_{{\rm f},k}}
\nwc{\RO}{\mathcal{R}_{\rm o}}
\nwc{\RF}{\mathcal{R}_{\rm f}}
\nwc{\ROk}{\mathcal{R}_{{\rm o},k}}
\nwc{\RFk}{\mathcal{R}_{{\rm f},k}}
\nwc{\QQ}{\mathcal{Q}}
\nwc{\PT}{\mathcal{T}}
\nwc{\real}{\text{re}}
\nwc{\imag}{\text{im}}
\nwc{\ep}{\epsilon}
\nwc{\lamb}{\lambda}
\nwc{\tlam}{{\lambda}_0}
\nwc{\hlam}{\hat{\lambda}}
\nwc{\tphi}{{{\phi}_0}}
\nwc{\CN}{\mathcal{C}(\cN )}

\nwc{\mf}{\mathbf}
\nwc{\mb}{\mathbf}
\nwc{\ml}{\mathcal}
\nwc{\bj}{{\mb j}}
\nwc{\bA}{{\mb \Phi}}
\nwc{\IA}{\mathbb{A}} 
\nwc{\bi}{\mathbf i}
\nwc{\bo}{\mathbf o}
\nwc{\IS}{\mathbb{S}}
\nwc{\IC}{\mathbb{C}} 
\nwc{\ID}{\mathbb{D}} 
\nwc{\IM}{\mathbb{M}} 
\nwc{\IP}{\mathbb{P}} 
\nwc{\bI}{\mathbf{I}} 
\nwc{\IE}{\mathbb{E}} 
\nwc{\IF}{\mathbb{F}} 
\nwc{\IG}{\mathbb{G}} 
\nwc{\IN}{\mathbb{N}} 
\nwc{\IQ}{\mathbb{Q}} 
\nwc{\IR}{\mathbb{R}} 
\nwc{\IT}{\mathbb{T}} 
\nwc{\IZ}{\mathbb{Z}} 
\nwc{\IV}{\mathbb{V}}
\nwc{\IX}{\mathbb{X}}
\nwc{\IY}{\mathbb{Y}}

\nwc{\cY}{{\ml Y}}
\nwc{\cP}{{\ml P}}
\nwc{\cQ}{{\ml Q}}
\nwc{\cL}{{\ml L}}
\nwc{\cX}{{\ml X}}
\nwc{\cW}{{\ml W}}
\nwc{\cZ}{{\ml Z}}
\nwc{\cR}{{\ml R}}
\nwc{\cV}{{\ml V}}
\nwc{\cT}{{\ml T}}
\nwc{\crV}{{\ml L}_{(\delta,\rho)}}
\nwc{\cC}{{\ml C}}
\nwc{\cA}{{\ml A}}
\nwc{\cK}{{\ml K}}
\nwc{\cB}{{\ml B}}
\nwc{\cD}{{\ml D}}
\nwc{\cF}{{\ml F}}
\nwc{\cS}{{\ml S}}
\nwc{\cM}{{\ml M}}
\nwc{\cG}{{\ml G}}
\nwc{\cH}{{\ml H}}

\nwc{\bT}{{\mb T}}
\nwc{\bM}{{\mb M}}
\nwc{\cbz}{\overline{\cB}_z}
\nwc{\supp}{{\hbox{\rm supp}}}
\nwc{\fR}{\mathfrak{R}}
\nwc{\bY}{\mathbf Y}

\nwc{\pft}{\cF^{-1}_2}
\nwc{\bU}{{\mb U}}
\nwc{\bPhi}{{\mb \Phi}}
\nwc{\bPsi}{{\mb \Psi}}
\nwc{\im}{{\rm i}}
\nwc{\bt}{\mathbf{t}}
\nwc{\bw}{{\mathbf w}}
\nwc{\mbm}{{\mathbf m}}
\nwc{\lbr}{\textlbrackdbl}
\nwc{\rbr}{\textrbrackdbl}
\nwc{\vzero}{{\mathbf 0}}
\nwc{\cN}{{\mathcal N}}
\nwc{\rbra}{\textrbrackdbl}
\nwc{\lbra}{\textlbrackdbl}
\nwc{\conv}{\hbox{conv}}
\nwc{\rank}{\hbox{rank}}

\nwc{\beq}{\begin{eqnarray}}
\nwc{\beqn}{\begin{eqnarray*}}
\nwc{\eeqn}{\end{eqnarray*}}
\nwc{\eeq}{\end{eqnarray}}
\nwc{\cle}{\preccurlyeq}
\nwc{\modpi}{{{\rm mod}\,2\pi}}
\nwc{\lb}{\llbracket}
\nwc{\rb}{\rrbracket}
\nwc{\cpo}{{\mathcal{P}_{\rm o}}}
\nwc{\cpm}{{\mathcal{P}_{\rm m}}}
\nwc{\ma}{\measuredangle}
\nwc{\om}{\omega}
\nwc{\z}{y}
\nwc{\lt}{\left}
\nwc{\rt}{\right}
\nwc{\half}{{1\over 2}}
\nwc{\diag}{\hbox{\rm diag}}
\begin{document}

\title{Fourier-Domain Fixed Point Algorithms with Coded Diffraction Patterns}
\author{Albert Fannjiang}
\address{Department of Mathematics, University of California, Davis, CA 95616}

\thanks{Research partially supported by NSF DMS and Simons Foundation Grant No. 275037}

\begin{abstract}
Fourier-domain Difference Map (FDM) for phase retrieval with two oversampled coded diffraction patterns  are proposed.
FDM is  a 3-parameter family of fixed point algorithms including Fourier-domain Hybrid-Projection-Reflection (FHPR) and Douglas-Rachford (FDR) algorithm. 
For generic complex objects without any object constraint, FDM yields a unique fixed point,  after proper projection back to the object domain, that is the true solution to
the phase retrieval problem up to a global phase factor.  
\end{abstract}


\maketitle 

\section{Introduction}

Fixed point algorithms  are among
the most effective algorithms for phase retrieval. These include Douglas-Rachford (DR) algorithm \cite{BCL02},
Hybrid-Projection-Reflection (HPR) algorithm  \cite{BCL03}  and the Difference Map (DM) \cite{Elser03},  all of which
are based on the projections onto the constraint sets, including  the object domain constraints
(positivity, support constraint etc)  and the Fourier magnitude constraint. 
Their performance is on a par with the industry standard such as the Hybrid-Input-Output (HIO) algorithm \cite{Fie} which is not of the pure projection type and notoriously hard to analyze \cite{BCL02,Mar}. 

The numerical challenge to any phasing algorithms is two-fold: 
the possibility  of multiple fixed points and the non-convexity of  the Fourier magnitude 
constraint. The latter is the nature of phase retrieval, independent of algorithms, while 
the former depends on the information content of the measured data as well as  the design of algorithm.

The behaviors of any fixed point algorithm 
depend on the ``landscape" of  the object domain. If there are multiple attractive fixed points, the iterations can stagnate; if there are multiple
hyperbolic fixed points, then a strange attractor may emerge  and
the iterations may exhibit a chaotic behavior.
In other words, 
the presence of multiple fixed point in the object domain
often severely deteriorate numerical performance, causing stagnation or even divergence   of the iterations. 

On the other hand, the presence of multiple fixed points in the Fourier domain may not be
a bad thing, as long as these fixed points correspond to the unique fixed point in the
object domain. On the contrary, the presence of multiple fixed points in the Fourier domain
is a form of relaxation and may help mitigate  the stagnation problem. 

Therefore uniqueness of the fixed point in the object domain is a first-order concern to the algorithm design just like uniqueness of phase retrieval solution is
to the measurement design. The latter, however,  is the prerequisite of the former. 

The purpose of the present work is to formulate the 3-parameter family of DM in the {\em Fourier
domain with two oversampled  coded diffraction patterns}, but without any object domain constraint,  and prove the uniqueness of fixed point after proper projection back to the object domain. 

The motivation for coded measurement  is to the uniqueness of
phase retrieval solution as established in \cite{unique} and 
the tremendous enhancement in numerical performance illustrated in \cite{rpi,pum}. 
\section{Coded diffraction patterns}
Let us first review
the set-up for  coded diffraction patterns.

 Let $f(\bn)$ be a discrete  object function with $\bn = (n_1,n_2,\cdots,n_d) \in \IZ^d$. 
Consider  the {object space} consisting  of all functions  supported in 
\[
  \cN= \{ 0\le n_1\le N_1, 0\le n_2\le N_2,\cdots, 0\leq n_d\leq N_d\}. 
  \]
We assume $d\geq 2$. 

With a {coherent illumination} under the Fraunhofer 
approximation, the free-space propagation between the object plane and the sensor  plane 
can be described by the Fourier transform \cite{BW} (with the proper coordinates and normalization). However, only the {\em intensities} of the Fourier transform
are measured on the sensor plane and constitute  the so called {\em diffraction pattern} given by 
 \beq
   \sum_{\bn =-\bN}^{\bN}\sum_{\mbm\in \cN} f(\mbm+\bn)\overline{f(\mbm)}
   e^{-\im 2\pi \bn\cdot \bom},\quad \bom=(w_1,\cdots,w_d)\in [0,1]^d,\quad \bN = (N_1,\cdots,N_d)\nn
   \eeq
   which is the Fourier transform of the autocorrelation
   \beqn
	  R_{f}(\bn)=\sum_{\mbm\in \cN}f(\mbm+\bn)\overline{f(\mbm)}.
	  \eeqn
Here and below the over-line notation means
complex conjugacy. 

Note that
$R_{f}$ is defined on the enlarged  grid
 \begin{equation*}
 \widetilde \cN = \{ (n_1,\cdots, n_d)\in \IZ^d: -N_1 \le n_1 \le N_1,\cdots, -N_d\le n_d\leq N_d \} 
 \end{equation*}
whose cardinality is roughly $2^d$ times that of $\cN$.
Hence by sampling  the diffraction pattern
 on the grid 
\beqn
\cL = \Big\{(w_1,\cdots,w_d)\ | \ w_j = 0,\frac{1}{2 N_j + 1},\frac{2}{2N_j + 1},\cdots,\frac{2N_j}{2N_j + 1}\Big\}
\eeqn
we can recover the autocorrelation function by the inverse Fourier transform. This is the {\em standard oversampling} with which  the diffraction pattern and the autocorrelation function become equivalent via the Fourier transform.  The remaining task is to 
recover $f$ from its autocorrelation function, the object domain constraints and the knowledge of $\mu$.


A coded diffraction pattern is measured with a mask
whose effect is multiplicative and results in  
a {\em masked object}  of the form $f(\bn) \mu(\bn)$ 
where $\{\mu(\bn)\}$ is an array of random variables representing the mask.   
In other words, a coded diffraction pattern is just the plain diffraction pattern of
a masked object. 

We will focus on the effect of {\em random phases} $\phi(\bn)$ in the mask function 
$
\mu(\bn)=|\mu|(\bn)e^{\im \phi(\bn)}
$
where  $\phi(\bn)$ are independent, continuous real-valued random variables and $|\mu|(\bn)\neq 0,\forall \bn\in \cL$ (i.e. the mask is transparent). 

\begin{figure}[h]
\centerline{\includegraphics[width=1\columnwidth]{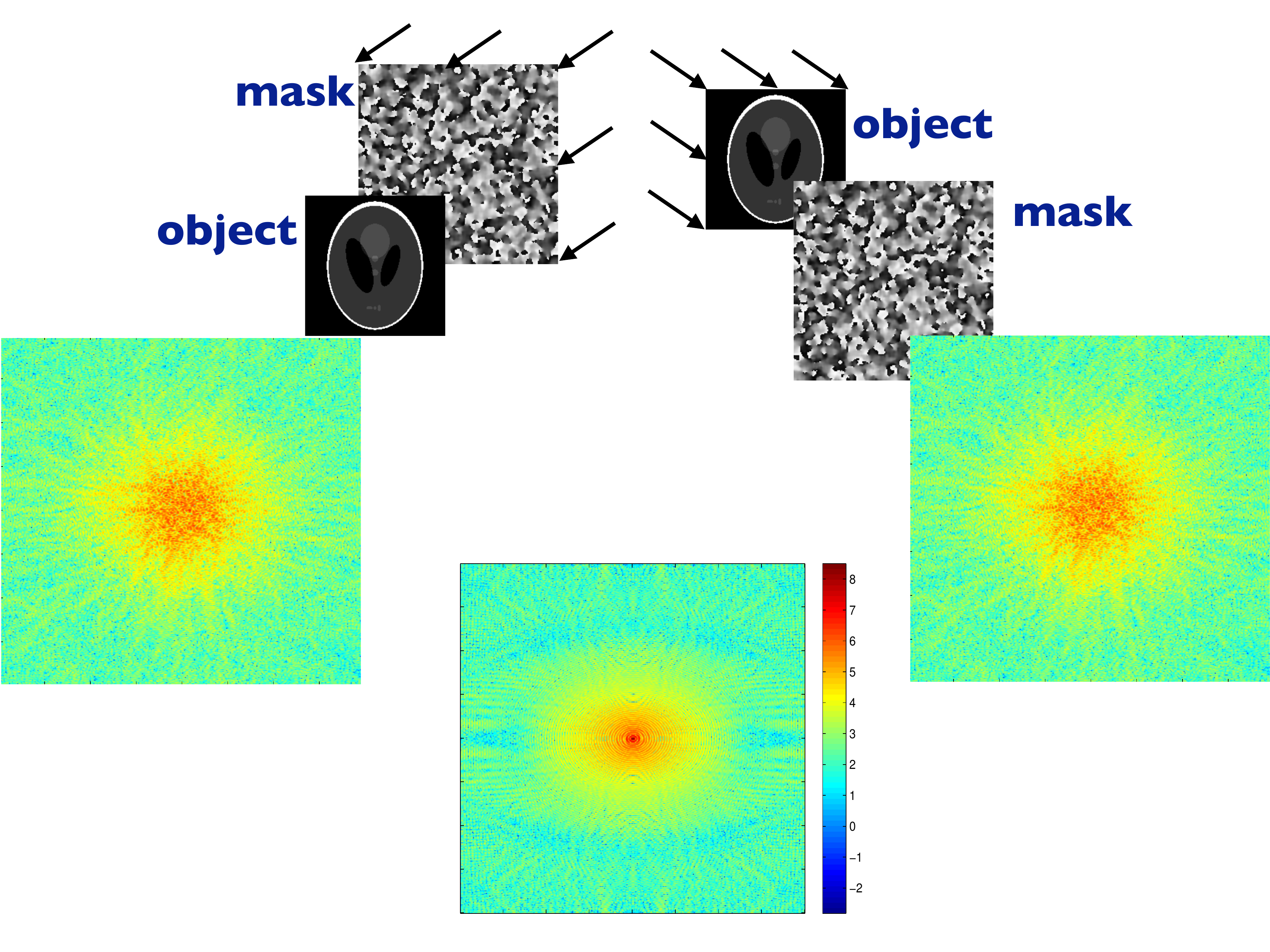}}
  \caption{Conceptual layout of coherent lensless imaging with a random mask (left) {\em before} (for random illumination) or (right) {\em behind} (for wavefront sensing) the object (phantom). 
 (middle) The diffraction pattern measured without a mask has a larger dynamic range. The color bar is on a logarithmic scale.} 
 \label{fig:mask}
\end{figure}

 Accordingly, let
$\Phi$ be the oversampled discrete Fourier transform from 
 $\cN$ to $\cL$ such that $\Phi^*\Phi=I$. In other words, $\Phi$ is an isometry and has orthonormal columns.  
  
 In the case of one masked measurement, the measurement matrix is $\Psi=\Phi~\diag(\mu) $ whereas in the case of two masked measurements, the measurement matrix is given by
\beq\Psi={1\over \sqrt{2}}\left[\begin{matrix}
\Psi_1\\
\Psi_2
\end{matrix}\right],\quad \Psi_j=\Phi~\diag(\mu_j),\quad j=1,2.
\label{0.0}
\eeq
where $\mu_1,  \mu_2$ are two independently generated masks.  Let $F=\Psi f$ be the mask-coded diffraction pattern(s). 

Now we recall the uniqueness of phase retrieval solution with two coded diffraction patterns \cite{unique}. 

\begin{proposition} \cite{unique}
\label{prop1}
Let $f$ be a complex-valued object
of dimension  $\geq 2$.   
Let $\Psi$ be the matrix given by \eqref{0.0}.  Let $g$ be   another complex object satisfying $|{\Psi f}| = |{\Psi g}|$ on $\cL$.
Then $g=e^{i\theta} f$, for some real constant $\theta$, with probability
one.
\end{proposition}

\section{Difference map in the Fourier domain}

For ease of presentation, we shall assume that the masks are {\em phase masks}, i.e. $|\mu_1(\bn)|=|\mu_2(\bn)|=1,\forall \bn$.  Consequently, $\Psi^*\Psi=I$. 

For ease of notation,  we convert the $d$-dimensional grid  into an ordered set of index.  For example, the unknown  object  $x_0\in \IC^{|\cN|}$ is the vectorized version of the object function $f$ originally supported in $\cN\subset \IZ^d, d\geq 2$.  
 
Let $ y\odot y'$ and $y/y'$ be the component-wise multiplication and division between  two vectors $y,y'$, respectively. For any $y\in \IC^{|\cL|}$ define the phase vector $\om\in \IC^{|\cL|}$ with  $\om(j)=\z(j)/|\z(j)|$ where $|\z(j)|\neq 0$.
When $|\z(j)|=0$ the phase can be assigned arbitrarily and we set $\om(j)=1$  unless otherwise specified.

Phase retrieval   can be formulated   as the following feasibility problem 
in the Fourier domain 
 \beq
 \label{feas}
\hbox{Find}\quad  \hat y\in  \Psi \cX \cap \cY,\quad \cY:=  \{y\in \IC^{|\cL|}: |y|=b\}.
\eeq
 
 
Let $\cpo$ be the projection onto $\Psi \cX$ and $\cpm$ the projection onto $\cY$:
\beq
\label{proj}
\cpo y=\Psi \Psi^* y, \quad \cpm y=b\odot{y\over |y|}
\eeq
The  Difference Map (DM) $\cD$ is defined as follows.
Let 
\beq
\cD=I+\beta \Delta\label{dm}
\eeq
with
\beq
\Delta&=&\cpo \big((1+\gamma_2)\cpm-\gamma_2 I\big)-\cpm\big((1+\gamma_1)\cpo-\gamma_1 I\big)\label{dm2}
\eeq
where $\beta\neq 0,\gamma_1,\gamma_2$ are three relaxation parameters. 

When $\gamma_1=-1$ and $\gamma_2=1/\beta$, \beq
\label{hpr}
\cD=I+\beta\Big(\cpo\big((1+{1\over \beta})\cpm-{1\over \beta}I\big)-\cpm\Big)
\eeq
FDM becomes FHPR which, with $\beta=1$, becomes FDR:
 \beq\label{dr}
S y &=&y+\Psi \Psi^*  \lt(2b\odot \frac{y}{|y|}-y\rt)-b\odot \frac{y}{|y|}. 
\label{fdr}
\eeq

\section{Uniqueness of fixed point}

FDM is so designed that its Fourier fixed points become the phase retrieval solution after
proper projection. 

Let $y_*$ be a fixed point of \eqref{dm} and hence satisfy
$\Delta y_*=0$, i.e.
\beq
\label{2}
\cpo \big((1+\gamma_2)\cpm-\gamma_2 I\big)y_*= \cpm\big((1+\gamma_1)\cpo-\gamma_1 I\big)y_*.
\eeq
Let
\beq
\label{3}v_*&\equiv&\big((1+\gamma_1)\cpo-\gamma_1I \big)y_*\\
\label{4}\eta_*&\equiv&\big((1+\gamma_2)\cpm-\gamma_2 I\big)y_*
\eeq
and 
\beq
\hat y\equiv \cpo \eta_*,\quad  \hat x\equiv \Psi^* \hat y=\Psi^*\eta_*. \label{4.1}
\eeq
By \eqref{2}  $\hat y=\cpm v_*=\cpo \eta_*$  and hence $\hat y$ satisfies both the object domain constraint represented by $\cpo$ as well as the Fourier domain constraint represented by $\cpm$. 

We now prove that 
DM produces the unique phase retrieval solution up to a constant phase factor.

\commentout{
\begin{remark}\label{rmk6.3}
With a slightly stronger assumption (\cite{unique}, Theorem 6),  the two-mask
case in Proposition \ref{prop6.1}  holds  for the
$1\half$-mask case \eqref{two'}.
\end{remark}
}

\begin{theorem} \label{thm4.1} Let $y_*$ be a fixed point of FDM and $v_*,\eta_*,\hat x, \hat y$ be defined by 
 \eqref{3}, \eqref{4} and \eqref{4.1}. Let $x_*=\Psi^*y_*$. The following statements hold
 with probability one. 
 \begin{itemize}
 \item[(i)]  $\hat y=e^{i\theta} y_0$ and $\hat x=e^{i\theta} x_0$  for some real constant $\theta$.
 
 \item[(ii)] If $\gamma_2\neq 0$ and $\gamma_1=-1$, then $\cpo y_*=e^{i\theta}y_0$ and 
  $x_*=e^{i\theta}x_0$,  for some real constant $\theta$. 
  
  \item[(iii)] If $\gamma_1=0$, then $y_*=e^{i\theta}y_0$  for some real constant $\theta$.
   
 \end{itemize}
\end{theorem}

\begin{remark}
Part (i) means  that in general $\hat y$, instead of $y_*$, is unique up to a constant phase factor. However, the relationship between $\hat y$ and $y_0$ is nonlinear. For example, $\eta_*$ and $y_*$ are already related pixel-wise via the complicated  relationship
\beq
|\eta_*|&=& \lt|(1+\gamma_2)b -\gamma_2|y_*|\rt|\\
\ma \eta_*&=&\ma y_*+\sigma \pi
\eeq
where $\sigma$ can take any of the three values $0,\pm \pi$ depending on the pixel.

In view of part (ii), on the other hand,  the relationship between $y_*$ and $y_0$ is linear and
the desirable property   $x_*=e^{i\theta}x_0$ holds for FHPR with $\beta\neq 0$. 

Part (iii) implies  uniqueness in the Fourier domain (as well as in the object domain) up to a global phase factor. 

\end{remark}
\begin{proof}

Eq. \eqref{2} implies that $\cpo \eta_* = \cpm v_*$,    namely 
$\cpo \eta_*$ shares the same {\em magnitude} as $ y_0$ and
the same {\em phase}  as  $v_*$ at every point in $ \cL$:
\beq
\big|\cpo \eta_*\big|&=& \big|y_0\big|\label{10}\\
\ma\cpo \eta_*&=& \ma v_* \label{11}
\eeq
on $\cL$.

By Proposition \ref{prop1} and \eqref{4.1},  \eqref{10} implies 
\beq
\label{10.1}
\hat y=e^{i\theta} y_0
\eeq
 for some real constant $\theta$ with  probability one.  This proves part (i).   

For part (ii), substituting \eqref{10.1} into \eqref{4}, we have
\beq
e^{i\theta} y_0&=&(1+\gamma_2) \cpo \cpm y_*-\gamma_2\cpo y_*.\label{10.3}
\eeq
On the other hand, \eqref{11} implies that 
\beq
\label{10.2}
\ma \hat y=\ma v_*=\ma y_*, 
\eeq
 under the assumption $\gamma_1=-1$, and hence $\cpm y_*=e^{i\theta} y_0$. 
 
 Now from \eqref{10.3}   it follows
 that
 \beqn
 e^{i\theta} y_0&=&e^{i\theta} (1+\gamma_2)\cpo y_0-\gamma_2 \cpo y_*\\
 &=&e^{i\theta} (1+\gamma_2) y_0-\gamma_2 \cpo y_*
 \eeqn
 and, since $\gamma_2\neq 0$, 
 \beq
 \label{10.5}
 \cpo y_*=e^{i\theta}y_0. 
 \eeq
 Applying $\Psi^*$ on the both sides of \eqref{10.5}, we obtain $x_*=e^{i\theta}x_0$.

\commentout{

As $y_*$ is a fixed point of $\cD$, 
\beq
\cpo \big((1+\gamma_2)\cpm-\gamma_2 I\big)y_*
&=&\cpm\big((1+\gamma_1)\cpo-\gamma_1 I\big)y_*.\label{2}
\eeq
Setting
\beq
\label{3}v_*&=&\big((1+\gamma_1)\cpo-\gamma_1I \big)y_*\\
\label{4}\eta_*&=&\big((1+\gamma_2)\cpm-\gamma_2 I\big)y_*
\eeq
we obtain that $\cpo \eta_*=\cpm v_*$,  namely 
$\cpo \eta_*$ shares the same {\em magnitude} as $ y_0$ and
the same {\em phase}  as  $v_*$ at every point in $ \cL$:
\beq
\big|\cpo \eta_*\big|&=& \big|y_0\big|\label{10}\\
\ma\cpo \eta_*&=& \ma v_* \label{11}
\eeq
on $\cL$.

By uniqueness of phase retrieval \eqref{10} implies 
\beq
\label{10.1}
\cpo \eta_*=e^{i\theta} y_0
\eeq
 for some real constant $\theta$ with  probability one.   

By uniqueness of magnitude retrieval (\ref{11}) implies that 
$ v_*=c\cpo \eta_*$ with a positive constant $c$.  Hence  we have $e^{i\theta} y_0=v_*/c$.

Substituting $v_*=ce^{i\theta} y_0$ into 
 (\ref{3}) gives
\beq
\label{5}
\gamma_1 y_*= (1+\gamma_1) \cpo y_* - {ce^{i\theta} y_0}.
\eeq
 Hence  $y_*=\cpo y_*$ implying  $y_*=ce^{i\theta} y_0$. 

We claim that $c=1$. This can be seen by substituting $y_*=ce^{i\theta} y_0$
into the fixed point equation \eqref{2} which becomes
\[
(1+\gamma_2) e^{i\theta} y_0-\gamma_2 c e^{i\theta} y_0= e^{i\theta} y_0
\]
implying $c=1$. 

\commentout{
Substituting $y_*=ce^{i\theta} y_0$ into (\ref{4}), we obtain
\beqn
\eta_*&=&(1+\gamma_2)\cpm (ce^{i\theta} {y_0})-c{\gamma_2} {e^{i\theta}y_0}\nn\\
&=&(1+\gamma_2)e^{i\theta} y_0- c{\gamma_2} {e^{i\theta}y_0}\nn\\
&=&(1+(1-c)\gamma_2)e^{i\theta} y_0.
\eeqn
By \eqref{10} 
\[
|(1+(1-c)\gamma_2)|=1 
\]
implying 
\[
c=1\quad \hbox{\rm or}\quad  1+2/\gamma_2.
\]
Therefore $y_*=e^{i\theta} y_0$ and $x_*=e^{i\theta}f$ as claimed. 
}

\commentout{
\subsection*{Two masks}
Writing  $y_0=[y_{1}, y_{2}]^T$ and $v_*=[v_{*1},v_{*2}]^T$, we have   
\[
e^{i\theta}c_jy_j=v_{*j}
\]
for some positive constant $c_j, j=1,2$, by \eqref{10.1} and uniqueness of magnitude retrieval.

Substituting $v_{*j}=e^{i\theta} y_j c_j$ into 
 (\ref{3}) gives
\beq
\label{5-2}
-\gamma_1 y_*= -(1+\gamma_1)\cpo y_*+e^{i\theta} \left[\begin{matrix}y_1 c_1\\
y_2 c_2
\end{matrix}\right].
\eeq
Note that 
\beq
\cpo\left[\begin{matrix}y_1 c_1\\
y_2 c_2
\end{matrix}\right]=
{1\over 2} (c_1+c_2)y_0.
\eeq
By applying $\cpo$ on both sides of (\ref{5-2}), we obtain
\beq
\label{21}
\cpo y_*={e^{i\theta} \over 2} (c_1+c_2)y_0
\eeq
and, after substituting it back in (\ref{5-2}),
\[
\gamma_1 y_*=e^{i\theta} {1+\gamma_1\over 2} (c_1+c_2)y_0
-e^{i\theta} \left[\begin{matrix}y_1 c_1\\
y_2 c_2
\end{matrix}\right].
\]
Hence
\beq
y_{*j}&=&e^{i\theta}  a_j y_j,\quad a_j= {(1+\gamma_1)(c_1+c_2)- 2c_j\over 2 \gamma_1}, \quad j=1,2.\label{22}
\eeq
Note that $a_j>0$, since $\gamma_1<-1$, and \[
a_1+a_2=c_1+c_2. 
\]

Applying $\Psi^* $ to (\ref{22}) we then have
\[
x_*={e^{i\theta}\over 2} (c_1+c_2)f.
\]

With (\ref{22}), the fixed point eq. (\ref{2}) becomes
\beq
{(1+\gamma_2)}y_0-{\gamma_2\over 2} (c_1+c_2)y_0
= \left[\begin{matrix}
\sigma_1 y_1\\
\sigma_2 y_2\end{matrix}\right]
\eeq
with
\beq
\sigma_j&=&\hbox{sgn}\left[-\gamma_1 a_j+{1+\gamma_1\over 2} (c_1+c_2)\right],\nn\\
&=&\hbox{sgn}(c_j)=1,\quad j=1,2.\nn
\eeq
\commentout{ We adopt the convention $\hbox{sgn}(0)=0$.  Note that 
\[
s\equiv \hbox{sgn}(a_1)+\hbox{sgn}{(a_2)}\in \{
2, 1, 0\}
\]
 since $a_1+a_2>0$. 
 }Since $\gamma_2\neq 0$, we have
 \beq
c_1+c_2&=&{2\over \gamma_2}(1+\gamma_2)-{2\over \gamma_2}=2. 
\eeq
Therefore $x_*=e^{i\theta} x_0$.
}
}

For part (iii), we also need  the uniqueness theorem of magnitude retrieval
which requires only one coded diffraction pattern. 
 \begin{proposition} \label{prop2} \cite{rpi,Hayes} 
 Let $x_0$ be a given rank $\geq 2$ object. If 
\beq
\label{mag}
\measuredangle \Psi \hat x=\measuredangle \Psi  x_0
\eeq
 (after proper
 adjustment of the angles wherever the coded diffraction patterns vanish), then almost surely $\hat x= c x_0$ for some positive constant $c$. 
\end{proposition}

With $\gamma_1=0$, $v_*=\cpo y_*$. 
By Proposition \ref{prop2},   (\ref{11}) implies that 
$ \Psi^*y_*=c\Psi^*\eta_*$ with a positive constant $c$.  Hence  from \eqref{10.1} we have $e^{i\theta} y_0=v_*/c$.

Substituting $v_*=ce^{i\theta} y_0$ into 
 (\ref{3}) gives
\beq
\label{5}
\gamma_1 y_*= (1+\gamma_1) \cpo y_* - {ce^{i\theta} y_0}.
\eeq
 Hence  $y_*=\cpo y_*$ implying  $y_*=ce^{i\theta} y_0$. 

We claim that $c=1$. This can be seen by substituting $y_*=ce^{i\theta} y_0$
into the fixed point equation \eqref{2} which becomes
\[
(1+\gamma_2) e^{i\theta} y_0-\gamma_2 c e^{i\theta} y_0= e^{i\theta} y_0
\]
implying $c=1$. \\
\end{proof}

{\bf Acknowledgements.} Research is supported in part by  US NSF grant DMS-1413373 and Simons Foundation grant 275037.\\

\end{document}